\documentclass[12pt,letterpaper]{article}          
\usepackage{osajnl}
\usepackage[draft]{hyperref} 
\newcommand{\bc}{\begin{center}}
\newcommand{\ec}{\end{center}}
\def\be{\begin{equation}}
\def\ee{\end{equation}}
\def\bea{\begin{eqnarray}}
\def\eea{\end{eqnarray}}

\def\la{\mathrel{\mathchoice {\vcenter{\offinterlineskip\halign{\hfil
$\displaystyle##$\hfil\cr<\cr\sim\cr}}}
{\vcenter{\offinterlineskip\halign{\hfil$\textstyle##$\hfil\cr
<\cr\sim\cr}}}
{\vcenter{\offinterlineskip\halign{\hfil$\scriptstyle##$\hfil\cr
<\cr\sim\cr}}}
{\vcenter{\offinterlineskip\halign{\hfil$\scriptscriptstyle##$\hfil\cr
<\cr\sim\cr}}}}}
\def\ga{\mathrel{\mathchoice {\vcenter{\offinterlineskip\halign{\hfil
$\displaystyle##$\hfil\cr>\cr\sim\cr}}}
{\vcenter{\offinterlineskip\halign{\hfil$\textstyle##$\hfil\cr
>\cr\sim\cr}}}
{\vcenter{\offinterlineskip\halign{\hfil$\scriptstyle##$\hfil\cr
>\cr\sim\cr}}}
{\vcenter{\offinterlineskip\halign{\hfil$\scriptscriptstyle##$\hfil\cr
>\cr\sim\cr}}}}}
\def\ve{\varepsilon}
\begin{document}

\title{Effective medium theories for irregular
fluffy structures:
aggregation of small particles}

\author{
Nikolai V. Voshchinnikov,
Gorden Videen,  and
Thomas Henning 
}

\address{

Sobolev
Astronomical Institute, St.~Petersburg University, St.~Petersburg, 198504 Russia \\
AMSRD--ARL--CI--ES, 2800 Powder Mill Road, Adelphi, Maryland 20783 USA \\
Max-Planck-Institut f\"ur Astronomie, K\"onigstuhl 17, D-69117 Heidelberg, Germany}

\email{nvv@astro.spbu.ru}

\begin{abstract}

We study the extinction efficiencies as well as
scattering properties of particles of different porosity.
Calculations are performed
for porous pseudospheres with small size (Rayleigh) inclusions
using the discrete dipole approximation.
Five refractive indices of materials covering the range from
$1.20+0.00i$ to $1.75+0.58i$
were selected. They
correspond to biological particles,
dirty ice, silicate, amorphous carbon and soot
in the visual part of spectrum.
We attempt to describe the optical properties of such particles
using Lorenz-Mie theory and a refractive index found from
some effective medium theory (EMT)
assuming the particle is homogeneous.
We refer to this as the effective model.

It is found that the deviations are minimal when utilizing
the EMT based on the Bruggeman mixing rule.
Usually the deviations in extinction factor do not 
exceed $\sim 5\%$ 
for particle porosity ${\cal P}=0 - 0.9$ and size parameters
$x_{\rm porous} = 2 \pi r_{\rm s,\,porous}/\lambda \la 25$.
The deviations are larger for scattering and absorption efficiencies
and smaller for particle albedo and asymmetry parameter.
Our calculations made for spheroids confirm these conclusions.
Preliminary consideration shows that the effective model represents
the intensity and polarization of radiation
scattered by fluffy aggregates quite well.
Thus, the effective models of spherical and non-spherical particles
can be used to significantly simplify computations
of the optical properties of aggregates containing
only Rayleigh inclusions.


\end{abstract}


\maketitle 

\section{Introduction}

Fluffy aggregate particles are encountered in the atmosphere and ocean,
interstellar clouds, and biological and chemical media. Finding their optical
properties is an important task for different fields of science and industry.
Great progress in the theoretical study of the light scattered by small
particles discerned in the last several years
makes it possible to calculate the optical properties of
arbitrary-shaped particles with anisotropic optical
properties and inclusions.\cite{mht00}
However, a major part of the numerical techniques
developed for aggregates  is still computationally intensive.
Moreover, the real structures of scatterers are poorly
known, making
detailed calculations often impossible.
Therefore, it is attractive to find a way to treat the optics
of large fluffy particles using simplified models;
for example, to replace the aggregates by some simplified
homogeneous particles with some average dielectric function.
(the approach is called the Effective Medium Theory EMT;
see Refs.~\onlinecite{BH83}, \onlinecite{cval00} for discussion).

There are many different mixing rules for dielectric functions
(see, e.g.,  Refs.  \onlinecite{sih99}--\onlinecite{maron05}).
They are rediscovered from time to time and sometimes
one effective medium expression can be derived from another one.
The EMTs for mixtures of materials are traditionally considered
in the framework of electrostatic fields.\cite{sih99}
Evidently, this restricts the range of applicability of the EMTs.
Note that previous considerations were given to small volume fractions
of inclusions in particles ($\la 20-40$\%).

In this paper, we consider particles consisting of vacuum and some material.
We analyze the optical properties of aggregate particles
using the discrete dipole approximation (DDA\cite{d00}) and compare them
with results using ``effective'' models; e.g., for porous pseudospheres, the 
scattering properties are determined assuming the sphere is homogeneous, 
and its refractive index is determined with an EMT.
The porosity is varied up to 90\% that corresponds to
very fluffy particles resembling aggregates with fractal dimension $<2$
{\rm (see, e.g., Ref.~\onlinecite{min06}).}

Some results for three-component composite particles
(silicate, carbon and vacuum) were already presented by
Voshchinnikov et al.\cite{vih04}
They show that the EMT approach can give rather
accurate results only if very porous particles have
so called ``Rayleigh'' inclusions
(small in comparison with the wavelength of incident radiation).
At the same time,  the optical properties of heterogeneous
spherical particles having inclusions of various sizes
(Rayleigh and non-Rayleigh) and very large porosity
are found to resemble  those of
spheres with a large number ($\ga 15-20$) of different layers.

The particle models are described in Sect.~\ref{mod}.
In Sect.~\ref{num_res} we present some illustrative results using
the effective model, size and refractive index of inclusions,
and particle shape variations.
Concluding remarks are given in Sect.~\ref{concl}.

\section{Models of Particles and Calculations}\label{mod}

We consider {\rm spherical}
particles consisting of some amount of a
material and some amount of vacuum. The amount of vacuum characterizes
the particle porosity ${\cal P}$ ($0 \leq {\cal P} < 1$), which
is introduced as
\be
{\cal P} = V_{\rm vac} /V_{\rm total}
= 1 - V_{\rm solid} /V_{\rm total}, \label{por}
\ee
where $V_{\rm vac}$ and $V_{\rm solid}$ are the volume fractions
of vacuum and solid material, respectively.
If ${\cal P}=0$  the particle is homogeneous and compact and its optical
properties are described by the Lorenz-Mie theory.
If the porosity is small we can consider the particle as a
solid matrix with vacuum inclusions.
If the porosity is large (the case of very fluffy aggregates)
the particle can be presented as a vacuum  ``matrix'' with solid inclusions.
For aggregates the porosity  can be represented as unity minus the volume
fraction of solid material in a sphere described around the aggregate.
Fluffy particles also can be presented as homogeneous
spheres of the same material mass
with a refractive index found using an EMT.
The size parameter of porous particles can be found as
\be
x_{\rm porous} = \frac{2 \pi r_{\rm s,\,porous}}{\lambda}
 = \frac{x_{\rm compact}}{(1-{\cal P})^{1/3}}
= \frac{x_{\rm compact}}{(V_{\rm solid} /V_{\rm total})^{1/3}}. \label{xpor}
\ee
So, $x_{\rm porous} = x_{\rm compact}$ if ${\cal P}=0$.

\subsection{DDA Calculations}

The calculations of the optical properties of particles with inclusions
are performed with the discrete dipole approximation (DDA).
We use the version  DDSCAT 6.0 developed by Draine \& Flatau\cite{df03}.
This technique can treat particles of arbitrary shape and
inhomogeneous structure.

The particles (``targets'' in the DDSCAT terminology) are constructed
employing
a special routine  producing quasispherical targets with
cubic inclusions of a fixed
size. The sizes of the target $d_{\rm max}$ and of the inclusions
$d_{\rm incl}$ are expressed in units of  the interdipole distance $d$.

In contrast to  previous modeling efforts
{\rm (e.g., Refs. \onlinecite{hs93}--\onlinecite{wcms-l94}),}
porous particles are not produced by removing dipoles
or inclusions from a target but  by attributing the refractive index
$m=1.000001+0.0i$ to the vacuum.

For the purpose of treating very porous particles, the number of dipoles
in the pseudospheres is taken to be quite large.
In all cases considered, particles with
maximum size $d_{\rm max}=91$ are studied.
 This value corresponds to the total number of dipoles in pseudospheres
$N_{\rm dip}=357128 - 381915$ depending on the
size of inclusions $d_{\rm incl}$.
 Thus, the criterion of the validity of the DDA for
 calculations of the extinction/scattering
cross sections $|m|kd < 1$
($m=n+ki$ is the complex refractive
index of the material, see Sect.~\ref{num_res} for its choice,
$k=2 \pi/\lambda$ the wavenumber with $\lambda$ being the wavelength
in vacuum)
of Draine \& Flatau\cite{df03} is satisfied up to size parameter
$x_{\rm porous} \approx 27 - 40$.

Targets {\rm with randomly distributed cubic inclusions}
with values of $d_{\rm incl}$  ranging from 1 to 5 are considered.
Note that the inclusions of the size $d_{\rm incl}=1$ are dipoles,
while the inclusions
with $d_{\rm incl}=3$ and 5 consist of 27 and 125 dipoles, respectively.
The optical characteristics of pseudospheres  with inclusions
are averaged over three
targets obtained for different random number sets. The calculations
show that in our case such an approach is practically equivalent to
time-consuming numerical averaging over target orientations.

\subsection{EMT Calculations}

An EMT allows one to determine
an effective dielectric function
$\varepsilon_{\rm eff}$
(the dielectric permittivity
is related to the refractive index as $\ve=m^2$)
of any heterogeneous particle consisting of several materials
with  dielectric functions $\varepsilon_i$.
EMTs are utilized extensively in optics of inhomogeneous media
{\rm (see discussion in Refs. \onlinecite{sih99},
\onlinecite{dw01}--\onlinecite{gue06} and
references therein).}
However, full systematic studies of the accuracy of different
mixing rules are lacking.

In this work we study several EMTs including the two
most often used, the Bruggeman and Garnett EMTs.
The formulas of mixing rules are collected in Table~\ref{emt} ($f$ is
the volume fraction of component ``1''),
and corresponding references can be found in Refs.~\onlinecite{maron05},
\onlinecite{v02}.
We usually consider that $f = V_{\rm solid} /V_{\rm total}$
and  $1-f = V_{\rm vac} /V_{\rm total}$.
Note also that the Garnett rule assumes that one material
is a matrix (host material) in which the other material is embedded.
When the roles of the inclusion and the host material are reversed,
the inverse Garnett rule is obtained.

\section{Numerical Results and Discussion}\label{num_res}
In this section, we present the results
illustrating the behaviour  of the efficiency factors or cross sections.
We consider primarily the extinction efficiency factor
$Q_{\rm ext}=C_{\rm ext}/\pi r_{\rm s}^2$, where
$C_{\rm ext}$ is the extinction cross section and
$r_{\rm s}$ the radius of spherical particle.
The refractive indices of compact particles are chosen to be
$m_{\rm compact} = 1.20+0.00i$,
$m_{\rm compact} = 1.33+0.01i$,
$m_{\rm compact} = 1.68+0.03i$,
$m_{\rm compact} = 1.98+0.23i$, and
$m_{\rm compact} = 1.75+0.58i$.
These values are typical of refractive indices of biological particles,
dirty ice, silicate, amorphous carbon and soot
in the visual part of the spectrum, respectively.
The refractive indices
 are taken from the Jena--Petersburg Database of
Optical Constants (JPDOC) described
in Refs.~\onlinecite{heal99}, \onlinecite{jetal02}
(for soot we used data published in Ref.~\onlinecite{chch90}).

\subsection{Effect of the size of inclusions}

\begin{figure}\bc
\resizebox{12cm}{!}{\includegraphics{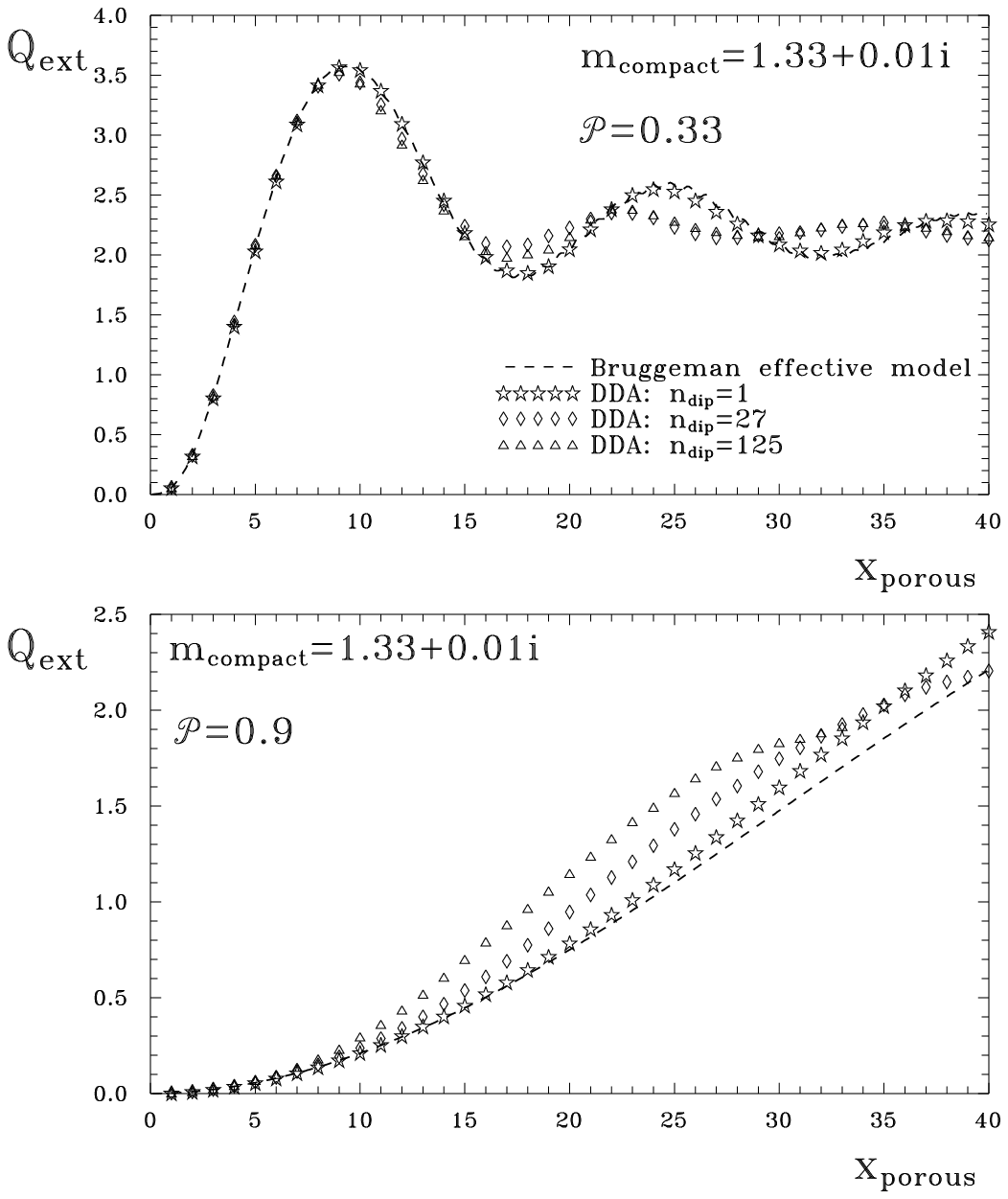}}
\caption{
}\label{ice}
\ec\end{figure}

The size of constituent particles (inclusions) is an important
parameter influencing light scattering by aggregates.
In Ref.~\onlinecite{vih04} it was demonstrated that
the Lorenz-Mie theory together with the standard EMTs (Garnett or Bruggeman)
reproduces the optical properties of aggregates
for particles with small (Rayleigh) inclusions only.
If the inclusions are not simple dipoles in the DDA terms,
the scattering characteristics of aggregates are not well reproduced
by the EMT calculations.
This fact is illustrated in Fig.~\ref{ice} where the
size dependence of the extinction efficiencies is plotted
for two values of particle porosity.
For illustration we choose the  Bruggeman EMT.

For particles consisting of cubes containing 27 and 125 dipoles,
the difference between the DDA and Bruggeman-EMT calculations becomes
quite large ($\ga 20\%$) for size parameters $x_{\rm porous} \ga 10$.
In this case the size parameter of the inclusions is 3  and 5 times larger
than for simple dipoles. This is enough to modify the pattern
of extinction. Larger inclusions produce curves having different slope
than simple dipoles and the Bruggeman-EMT. This conclusion is
valid for other factors and other refractive indices.

Note that the mixing rules with non-Rayleigh inclusions were developed in the
context of the extended EMT theory
(see, for example, the discussion in Ref.~\onlinecite{cval00}).
For aggregates consisting of inclusions of various sizes
(Rayleigh and  non-Rayleigh), a model of layered particles
can be applied
(see discussion in Ref.~\onlinecite{vih04}).
{\rm Below we consider particles with simple dipole inclusions
only.}

\subsection{Choice of the EMT}

Figure~\ref{icexx} shows the normalized extinction cross sections
$C_{\rm ext}^{\rm (n)}$ for aggregates with small (Rayleigh)
inclusions and the effective models based on the Lorenz-Mie calculations
with five different EMTs.
The normalized cross sections are calculated as
\be
C^{\rm (n)} = \frac{C({\rm porous \, particle})}
{C({\rm  compact \, particle \, of \, same \, mass})} =
\,\,\,\,\,\,\, (1-{\cal P})^{-2/3}\,  \frac{Q({\rm porous \, particle})}
{Q({\rm  compact \, particle \, of \, same \, mass})}. \label{cn}
\ee
They allow one to analyze the role of porosity in particle optics.
The quantity $C^{\rm (n)}$ shows how porosity increases or decreases
the cross section.
Three panels in Fig.~\ref{icexx} provide results for particles of
different masses. For each panel the mass of the particle remains
constant but its size increases according to Eq.~(\ref{xpor}).
The refractive index of compact particles is equal to
$m_{\rm compact} = 1.330+0.010i$.
The refractive indices of porous particles generally decrease
with the growth of porosity. Their values are given in Table~\ref{t-ref}
for three values of ${\cal P}$.
As follows from Table~\ref{t-ref} the difference between the values of $m$
is not large, but is enough to produce a
noticeable difference of the extinction efficiencies especially at large
porosity (see Fig.~\ref{icexx}).
The largest and smallest values of the effective refractive indices
(both real and imaginary parts) are obtained from the Birchak and Lichtenecker
mixing rules, respectively. Correspondingly, the properties calculated
with these $m$s deviate most strongly from the properties for
aggregates. We also find the relative deviations in the efficiency
factors (in percents) as
\begin{equation}
{\rm Deviation} = \frac{Q({\rm \mbox{EMT-Mie}}) - Q({\rm DDA})}
{Q({\rm DDA})} \; {\cdot} \; 100 \%. \label{err}
\end{equation}
Note that the deviations for particles of different mass and porosity
$\la 5\%$ if the  Bruggeman, Garnett or Looyenga mixing rule is used.
However, the deviation becomes $> 5\%$ for the Birchak and Lichtenecker rules
(see Fig.~\ref{iceerr}).
From Fig.~\ref{iceerr} it is seen that the effective models based on
the  Bruggeman and Looyenga rules reproduce the extinction
of aggregates (deviation $\la 1\%$) rather well if ${\cal P} \la 0.7$.
For larger porosity the Bruggeman model works better.
The usage of the Garnett rule leads to deviations within $\sim 4\%$
yielding properties generally smaller than those for aggregates.

Our calculations made for other mixing rules (e.g., quasi-crystalline,
coherent potential, see expressions in Refs.~\onlinecite{kolgust01, v02})
show that these rules
cannot reproduce even the general behaviour of the extinction
(e.g., $C_{\rm ext}^{\rm (n)}$ increase with the growth
of porosity for $x_{\rm compact} = 1$).
Based on the data presented in Figs.~\ref{icexx} and \ref{iceerr},
the three best effective models
(with  Bruggeman, Garnett and Looyenga mixing rules) are chosen for further
analysis.
The results for these three models and aggregates are shown in
Fig.~\ref{x3cs}.
This Figure is plotted for one value of $x_{\rm compact} = 3$ and
two values of $m_{\rm compact}$ corresponding to silicate and carbon in
the visible part of spectrum.
It is seen that the best results are obtained if the model based
on the Bruggeman rule is applied. It provides extinctions resembling those of
aggregates with small inclusions for particles of different size parameters,
porosity and refractive indices of inclusions.
So, further considerations are made on the models with the Bruggeman rule.

\subsection{Effect of the refractive index of inclusions}

The discussion above is {\rm mainly}
relevant to porous water ice in the visible part of spectrum.
Now we consider particles with inclusions of different refractive indices.
The comparison between the DDA and Bruggeman calculations is made
in Figs.~\ref{033} and \ref{09} for two particle porosity
${\cal P}=0.33$ and 0.9.
The upper panels show the extinction efficiency factors dependence
on the size parameter $x_{\rm porous}$. Five different refractive indices
have been considered. The effective refractive indices
found with the Bruggeman rule are indicated in Table~\ref{t-ref1}.
It is seen that the effective models describe
the general behaviour of extinction rather well. In all cases
the deviations between the factors $Q_{\rm ext}$ found for the aggregate
and for the effective model do not exceed $\sim 5\%$
(see lower panels of Figs.~\ref{033} and \ref{09}).
The exception is the case of silicate particles
($m_{\rm compact} = 1.680+0.030i$) and the porosity ${\cal P}=0.33$.
The Lorenz-Mie theory produces the ripple structure of the extinction for these
particles. Such structure does not appear in our DDA calculations.
This is because our targets are not smooth spheres but
pseudospheres whose cubic inclusions effectively destroy the resonances.

\subsection{Other factors}

We also consider how well the effective model reproduces
the scattering ($Q_{\rm sca}$) and absorption ($Q_{\rm abs}$) efficiencies,
the particle albedo
       \be\label{alb}
\Lambda =  \frac{Q_{\rm sca}}{Q_{\rm ext}}
       \ee
and the asymmetry parameter of the phase function $F(\Theta, \Phi)$
       \be\label{g}
       g = \langle \cos \Theta \rangle = \frac
       {\int_{4 \pi} F(\Theta, \Phi)\, \cos \Theta \, {\rm d}\omega}
       {\int_{4 \pi} F(\Theta, \Phi)\, {\rm d}\omega}.
       \ee
These quantities are plotted in Fig.~\ref{all}.
The comparison is made for the refractive indices of inclusions
$m_{\rm compact} = 1.33+0.01i$ and particle porosity ${\cal P}=0.9$.
It is seen that the agreement of results of the DDA
and the Bruggeman--Mie computations is rather good.
Our calculations performed for other values of ${\cal P}$
and $m_{\rm compact}$ show that the effective models better
reproduce the extinction properties than the scattering and absorption
properties. In the latter case the relative deviation usually does not exceed 10\%
(in comparison with 5\% for extinction).
At the same time albedo and the asymmetry parameter are reproduced
by the effective models with high accuracy: the relative deviation
usually does not exceed 2\%.

\subsection{Effect of the particle shape}

All previous results have been obtained for fluffy spherical particles
that can serve as an approximate model of aggregate particles 
randomly oriented in space
(3D orientation). If the aggregates have
a preferential axis of rotation (2D orientation) they can be considered as
fluffy axisymmetric particles (e.g., prolate or oblate spheroids).

We perform DDA calculations of the efficiency factors for
targets having the shape of prolate spheroids with Rayleigh inclusions.
The results are compared with those calculations performed using the
separation of variables methods (SVM, see Ref.~\onlinecite{vf93})
for homogeneous spheroids whose effective refractive index
is found from the Bruggeman EMT. Figure~\ref{sph} shows
the size dependence of the extinction efficiencies for prolate
spheroids with aspect ratio $a/b=2$ for the case of the incident radiation
propagating along the rotation axis of the spheroid ($\alpha=0^0$).
Note that for the considered case, the agreement between
the DDA and the Bruggeman--SVM computations is even better than
for spheres (cf. Fig.~\ref{09}):
the relative deviations $\la 4\%$ for $x_{\rm porous} \la 40$.
So, the effective models of non-spherical particles
seems to improve the accuracy of the effective model computations
for aggregates containing small size inclusions.

\subsection{Intensity and polarization}\label{int-p}

We also perform illustrative calculations of the intensity and
polarization of scattered radiation (see Fig.~\ref{ip}).
It is seen that satisfactory agreement between the effective model
and DDA computations is obtained for small and intermediate scattering
angles ($\Theta \la 60^0$) only. For larger scattering angles the
difference becomes rather large, especially for the second and third minima.
This is not unexpected, since diffraction plays a major role for 
small scattering angles, and this depends primarily on the external
morphology of the particle. At larger scattering angles, the internal 
composition plays a larger role. However, the deviations in reproducing
these minima are a
small concern when we consider a natural polydispersion of particles.
In this case, the minima become washed out due to the polydispersion.

\section{Conclusions}\label{concl}

We study the general optical behaviour of aggregate particles
when the porosity increases.

The main results of the paper are the following:

1. Extinction produced by porous pseudospheres with small size
(Rayleigh) inclusions can be calculated employing
the Lorenz-Mie theory with the refractive index found using 
an EMT. The deviations that arise using the Bruggeman effective model
do not exceed $\sim 5\%$ for particle porosity ${\cal P}=0 - 0.9$ and
size parameters $x_{\rm porous} \la 25$.

2.  The effective models represent the behaviour of other properties
(scattering and absorption efficiencies, particle albedo,
asymmetry parameter) quite well and can be used for calculations of
the intensity and polarization of radiation scattered by fluffy
aggregates under certain conditions. Preliminary consideration 
shows that the above
conclusions are also valid for spheroidal particles.

3. The effective models can significantly simplify computations
of the optical properties of aggregates containing only Rayleigh
inclusions.

\section{Acknowledgments}
We thank Vladimir Il'in and anonymous reviewers
for helpful comments.
We are grateful to Bruce Draine and Piotr Flatau for providing
DDSCAT 6.0 code.
The work was partly supported by
the TechBase Program on Chemical and Biological Defense,
by the Battlefield Environment Directorate under the auspices
of the U.S. Army Research Office Scientific Services Program administrated
by Batelle (Delivery Order 0395, Contract No. DAAD19-02-D-0001)
and  by grant of the DFG Research Group ``Laboratory Astrophysics''
and by grants NSh 8542.2006.2, RNP 2.1.1.2152 and RFBR 07-02-0000
of the Russian Federation.




\newpage

\begin{table}
\bc
\caption{Mixing rules for the refractive indices.
}
\label{emt}
\begin{tabular}{lc}
Mixing rule & Formula \\
\noalign{\smallskip}\hline\noalign{\medskip}
Bruggeman  &
$
\displaystyle f \frac{\ve_1 - \ve_{\rm eff}}{\ve_1 + 2 \ve_{\rm eff}} +
(1-f) \frac{\ve_2 - \ve_{\rm eff}}{\ve_2 + 2 \ve_{\rm eff}} =0
\label{Bru}
$    \\            \noalign{\medskip}
Garnett  &
$
\displaystyle\ve_{\rm eff} = \ve_2 \left[1 +
\frac{ 3f \frac{\ve_{1} - \ve_{\rm 2}}{\ve_{1} + 2 \ve_{\rm 2}}}
{ 1 - f \frac{\ve_{1} - \ve_{\rm 2}}{\ve_{1} + 2 \ve_{\rm 2}}}
\right]
\label{Gar}
$  \\         \noalign{\medskip}
Inverse Garnett &
$\displaystyle\ve_{\rm eff} = \ve_1 \left[1 +
\frac{ 3 (1-f) \frac{\ve_{2} - \ve_{\rm 1}}{\ve_{2} + 2 \ve_{\rm 1}}}
{ 1 - (1-f) \frac{\ve_{2} - \ve_{\rm 1}}{\ve_{2} + 2 \ve_{\rm 1}}}
\right]
$       \\         \noalign{\medskip}
Looyenga    &
$
\ve_{\rm eff}^{1/3}=  f \ve_1^{1/3} + (1-f)\ve_2^{1/3}
\label{Loo}
$          \\      \noalign{\medskip}
Birchak        &
$
\ve_{\rm eff}^{1/2}=  f \ve_1^{1/2} + (1-f)\ve_2^{1/2}
\label{Bir}
$             \\   \noalign{\medskip}
Lichtenecker        &
$
\log \ve_{\rm eff} = f \log \ve_1 + (1-f) \log \ve_2
\label{Licht}
$                \\
\noalign{\medskip}\hline\noalign{\smallskip}
\end{tabular}
\ec
\end{table}

\newpage

\begin{table}
\bc
\caption{Effective refractive indices $m=n+ki$
of porous particles  calculated using different EMTs
presented in Fig.~\ref{icexx} ($m_{\rm compact} = 1.330+0.010i$).}
\label{t-ref}
\begin{tabular}{lccc}
\hline\noalign{\smallskip}
Mixing rule & \multicolumn{3}{c}{Porosity}  \\
\noalign{\smallskip}  \cline{2-4}\noalign{\smallskip}
 & ${\cal P}=0.3$ & ${\cal P}=0.5$ &  ${\cal P}=0.9$ \\
\noalign{\smallskip}\hline\noalign{\smallskip}
Bruggeman   & $1.2284+0.0069i$   & $1.1611+0.0048i$  & $1.0310+0.0009i$ \\
Garnett     & $1.2247+0.0066i$   & $1.1579+0.0045i$  & $1.0308+0.0008i$  \\
Looyenga    & $1.2277+0.0068i$   & $1.1611+0.0048i$  & $1.0316+0.0009i$  \\
Birchak     & $1.2310+0.0070i$   & $1.1650+0.0050i$  & $1.0330+0.0010i$  \\
Lichtenecker& $1.2210+0.0064i$   & $1.1533+0.0043i$  & $1.0289+0.0008i$ \\
\noalign{\smallskip}\hline
\end{tabular}
\ec
\end{table}

\newpage

\begin{table}
\bc
\caption{Effective refractive indices $m=n+ki$
of porous particles  calculated using the Bruggeman EMT
presented in Figs.~\ref{033} and \ref{09}.}
\label{t-ref1}
\begin{tabular}{ccc}
\hline\noalign{\smallskip}
${\cal P}=0$ & ${\cal P}=0.33$ &  ${\cal P}=0.9$ \\
\noalign{\smallskip}\hline\noalign{\smallskip}
$1.2000+0.0000i$   & $1.1328+0.0000i$ & $1.0193+0.0000i$ \\
$1.3300+0.0100i$   & $1.2183+0.0066i$ & $1.0310+0.0009i$ \\
$1.6800+0.0300i$   & $1.4471+0.0196i$ & $1.0588+0.0022i$ \\
$1.9800+0.2300i$   & $1.6431+0.1507i$ & $1.0795+0.0137i$ \\
$1.7500+0.5800i$   & $1.4916+0.3781i$ & $1.0707+0.03932$ \\
\noalign{\smallskip}\hline
\end{tabular}
\ec
\end{table}

\newpage

\newpage

\begin{figure}\bc
\resizebox{10cm}{!}{\includegraphics{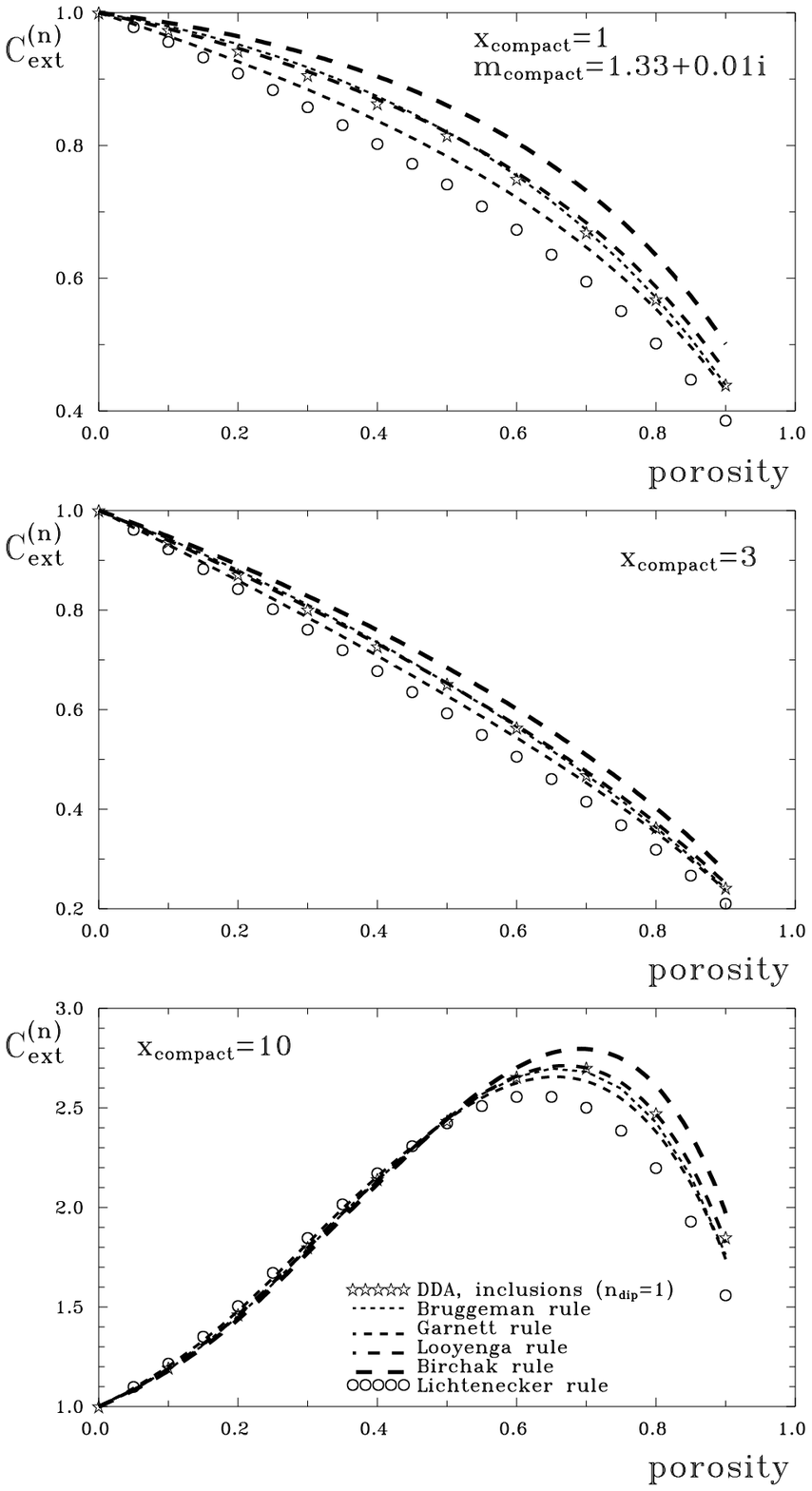}}
\caption{
}\label{icexx}
\ec\end{figure}

\newpage

\begin{figure}\bc
\resizebox{12cm}{!}{\includegraphics{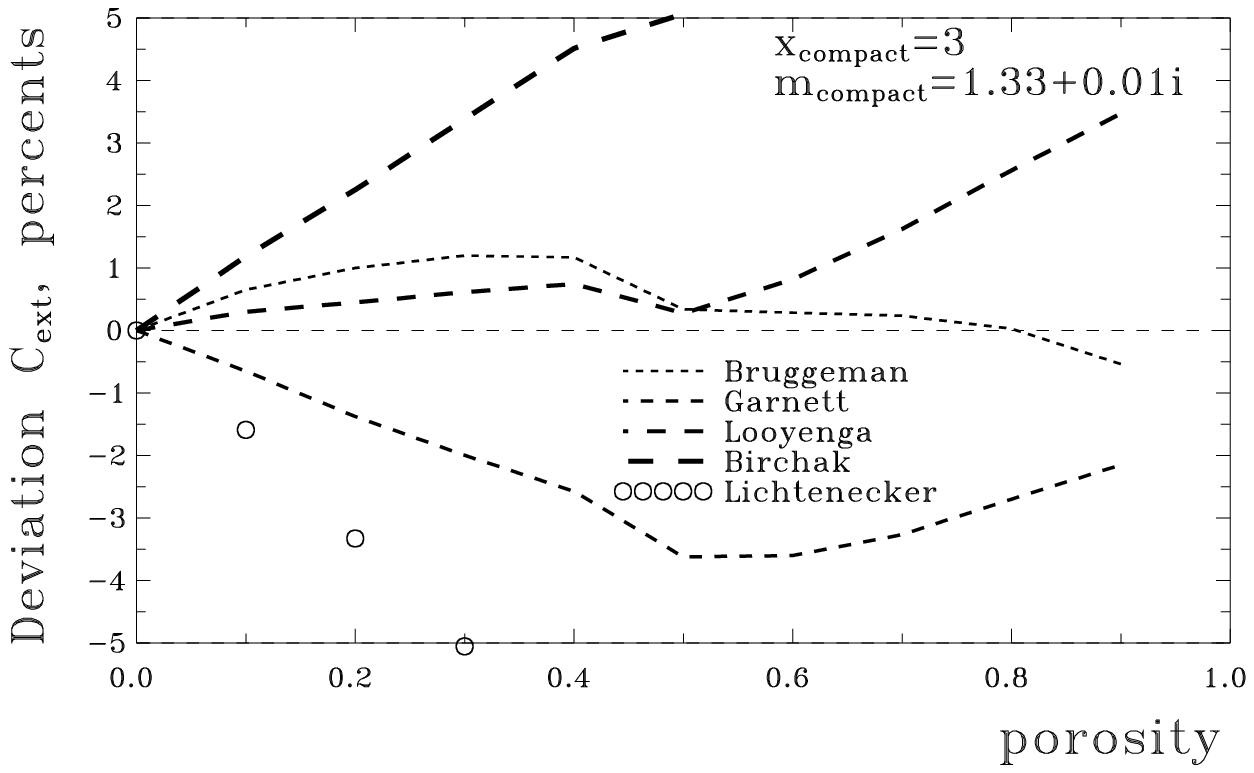}}
\caption{
}\label{iceerr}
\ec\end{figure}

\newpage

\begin{figure}\bc
\resizebox{12cm}{!}{\includegraphics{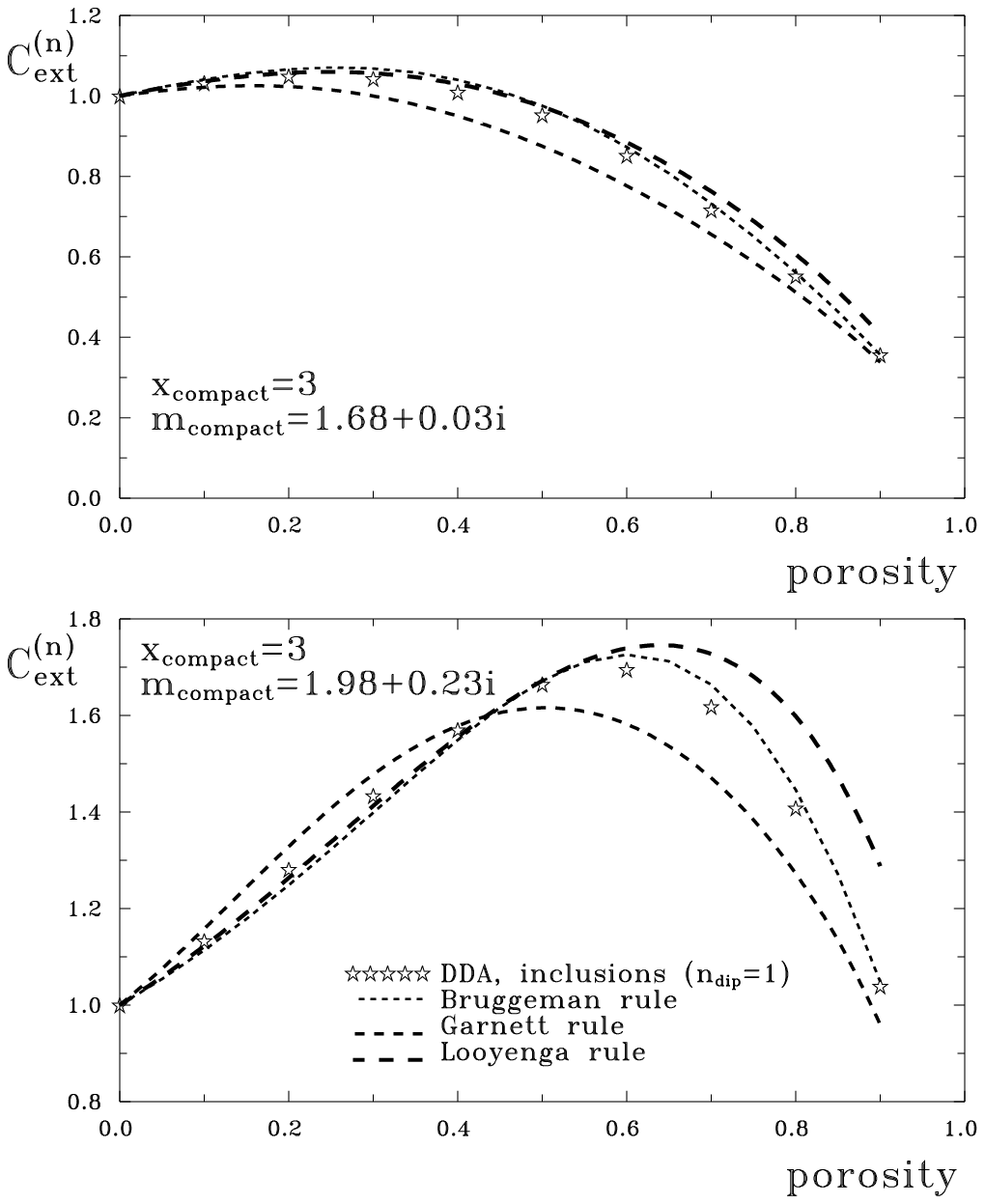}}
\caption{
}\label{x3cs}
\ec\end{figure}

\newpage

\begin{figure}\bc
\resizebox{12cm}{!}{\includegraphics{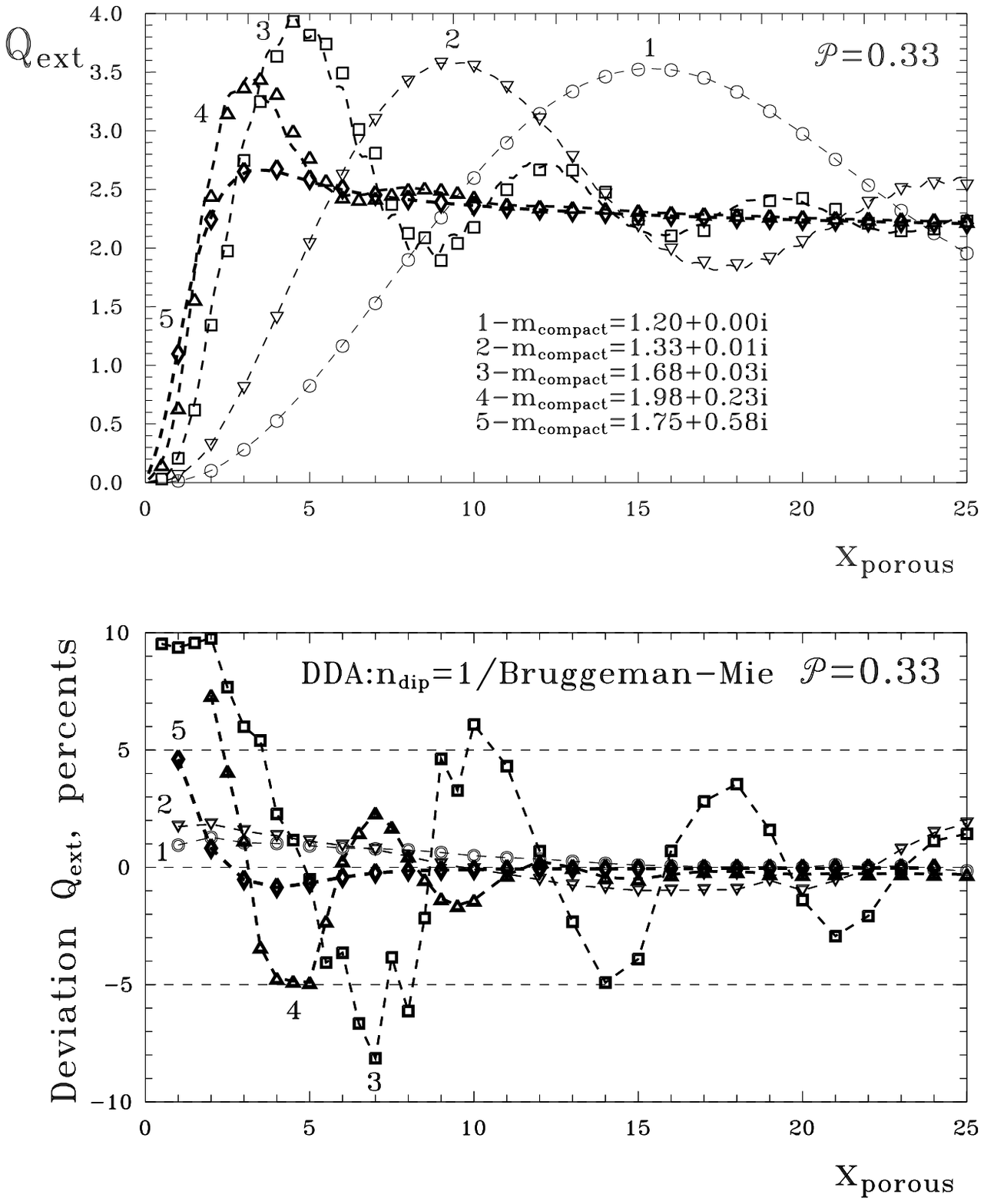}}
\caption{
}\label{033}
\ec\end{figure}

\newpage
\begin{figure}\bc
\resizebox{12cm}{!}{\includegraphics{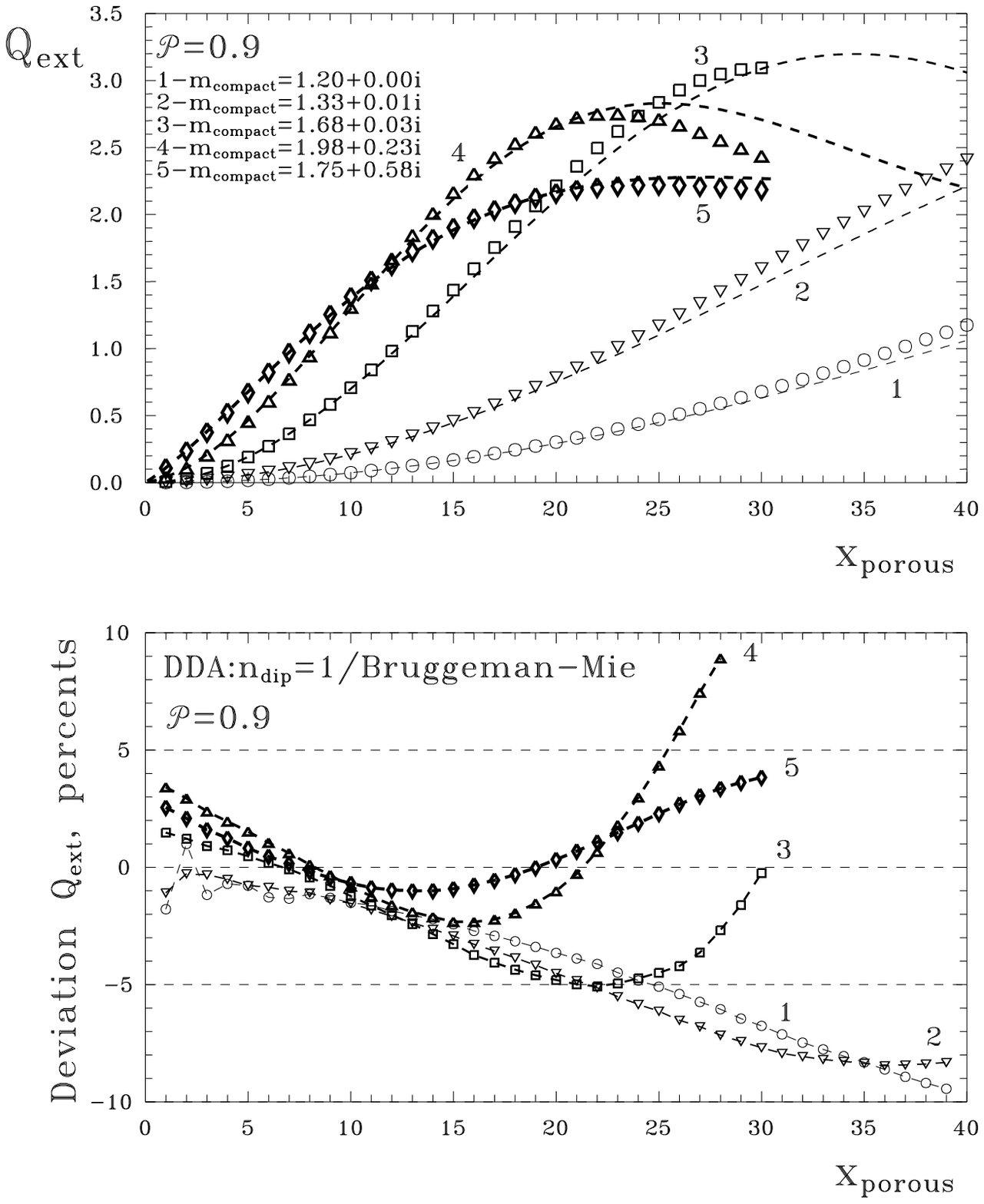}}
\caption{
}\label{09}
\ec\end{figure}

\newpage

\begin{figure}\bc
\resizebox{\hsize}{!}{\includegraphics{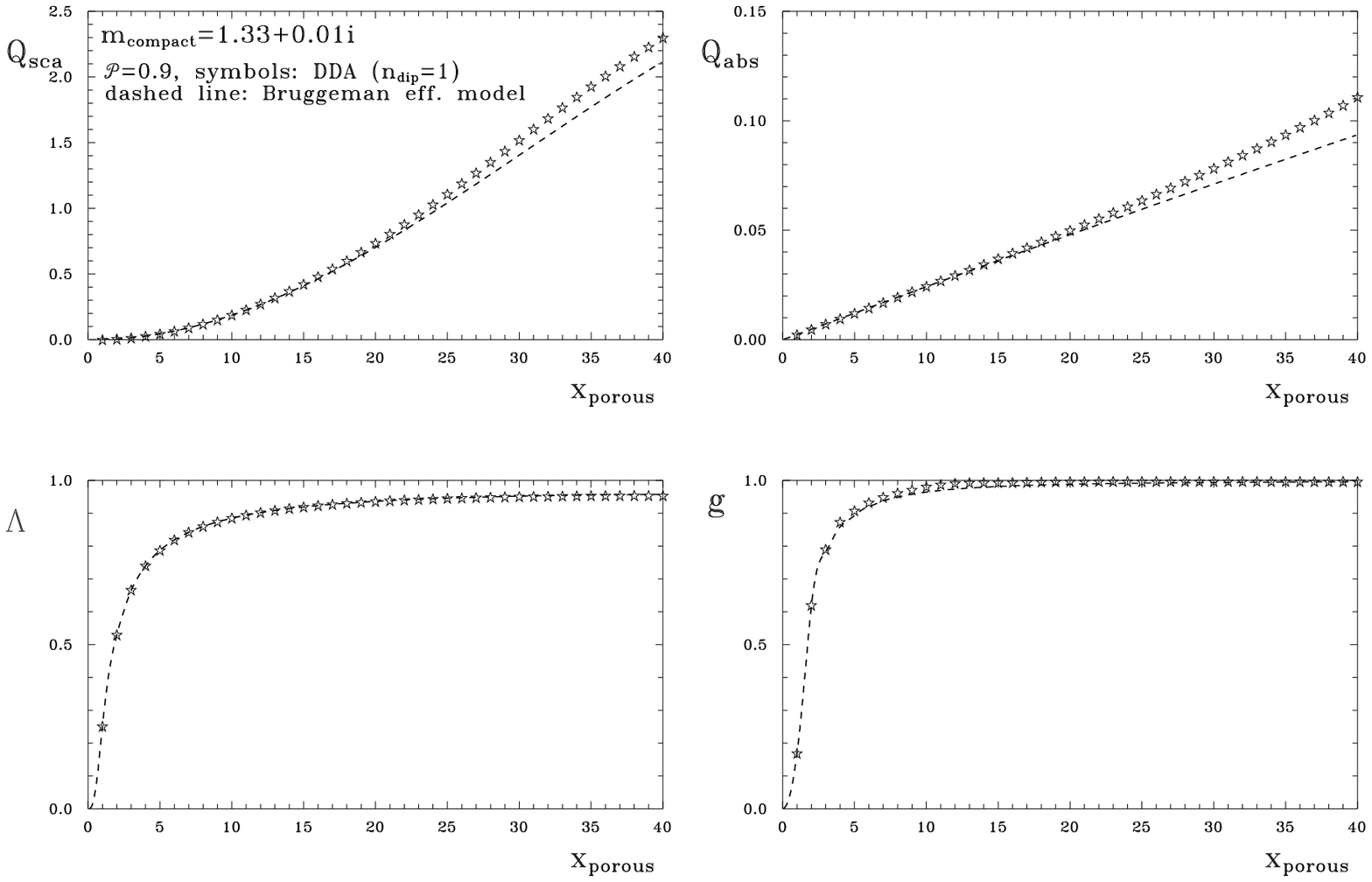}}
\caption{
}\label{all}
\ec\end{figure}

\newpage

\begin{figure}\bc
\resizebox{12cm}{!}{\includegraphics{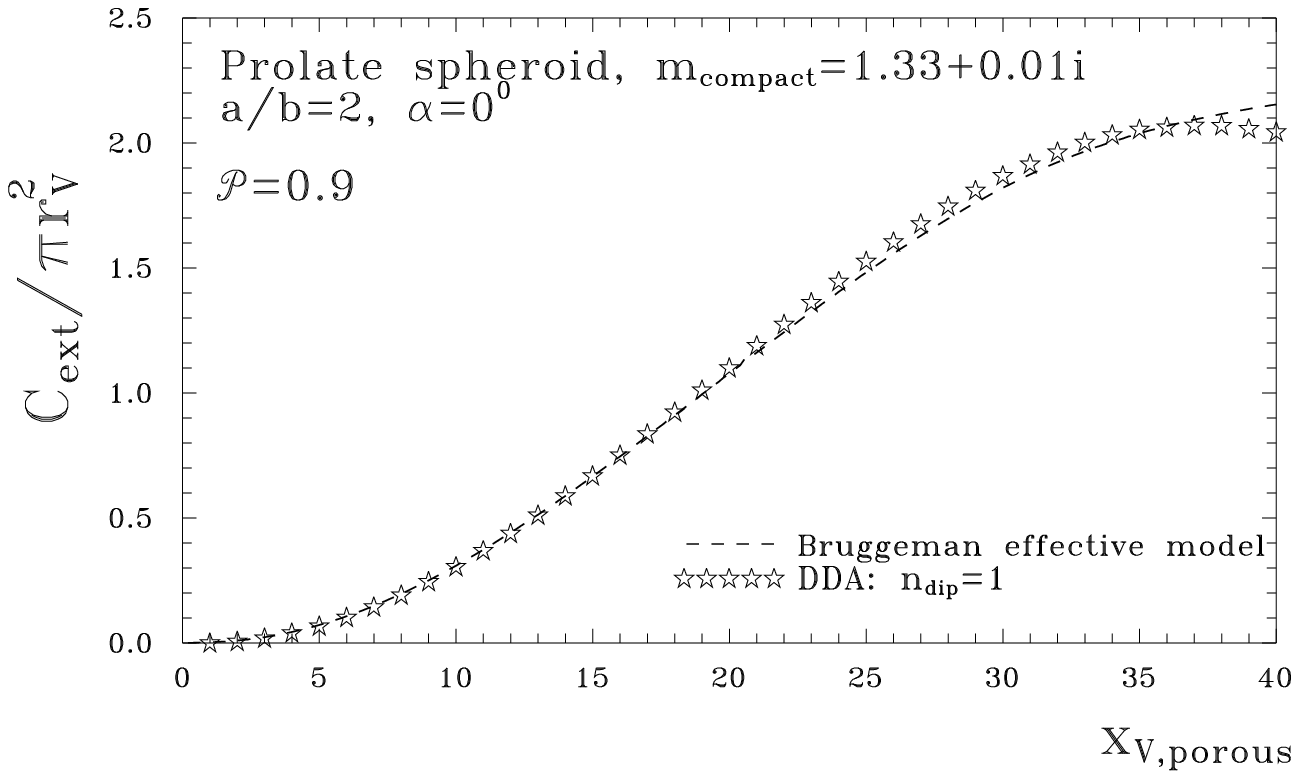}}
\caption{
}\label{sph}
\ec\end{figure}

\newpage

\begin{figure}\bc
\resizebox{12cm}{!}{\includegraphics{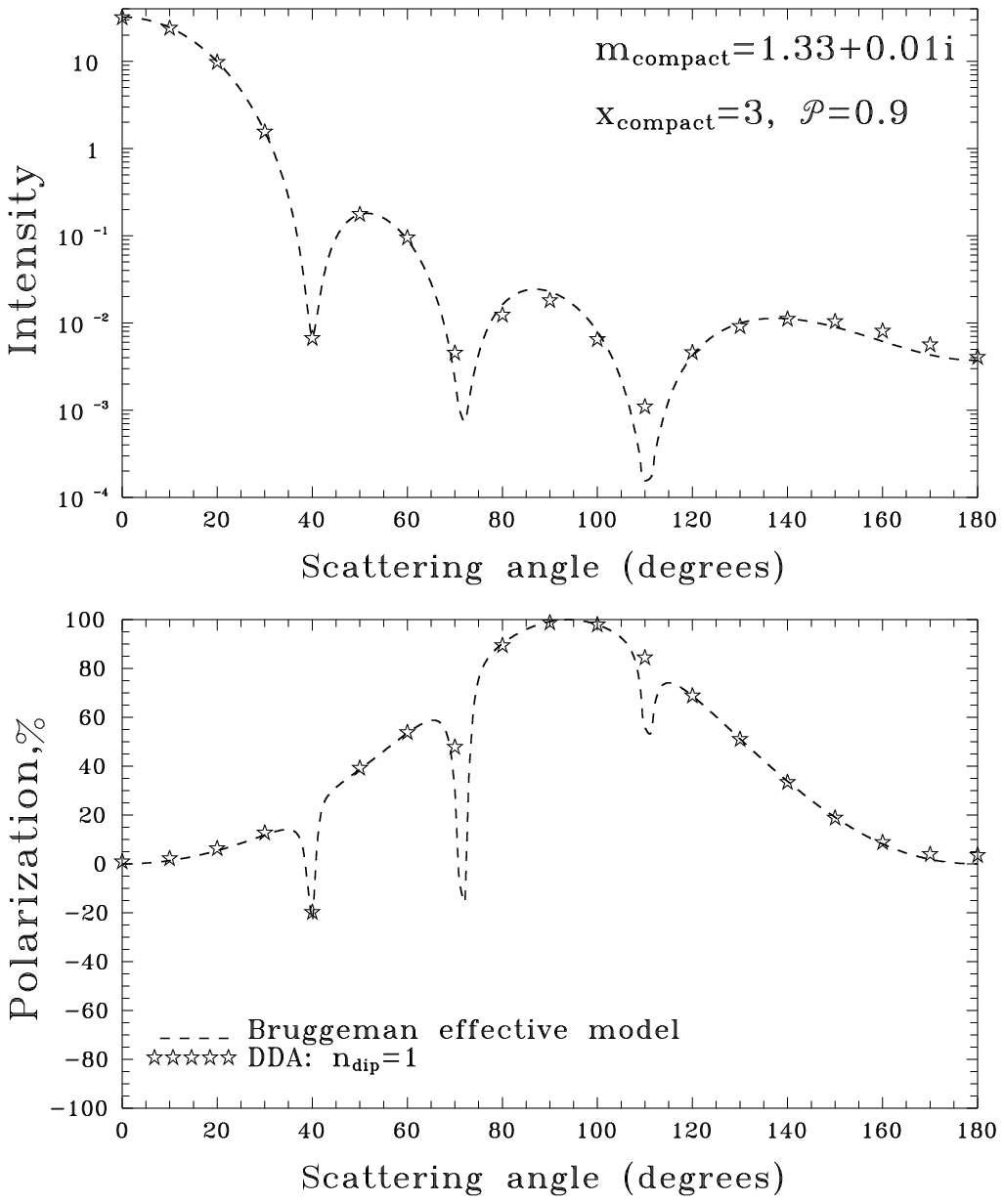}}
\caption{
}\label{ip}
\ec\end{figure}

\newpage
\clearpage

\bc FIGURE CAPTIONS \ec

\noindent Fig. 1.
   Size dependence of the extinction efficiency factors
   calculated for spheres with inclusions of different sizes
   (DDA computations) and with the Lorenz-Mie theory using the Bruggeman 
EMT.
   The refractive index of inclusions is $m_{\rm compact} = 1.33+0.01i$.
   The effective refractive indices of porous particles are indicated in Table~\ref{t-ref1}.
   The porosity of particles is ${\cal P}=0.33$ (upper panel) and
   ${\cal P}=0.9$ (lower panel).
   For given porosity the particles of the same size parameter $x_{\rm porous}$
   have the same mass.
The effect of variations of the size of inclusions is illustrated.

\bigskip

\noindent Fig. 2.
   Porosity dependence of the normalized extinction cross sections
   calculated for spheres with small inclusions
   (DDA computations) and with the Lorenz-Mie theory using different EMTs
($m_{\rm compact} = 1.33+0.010i$).
The effects  of variations of the EMT and particle size are illustrated.

\bigskip

\noindent Fig. 3.
   Dependence of the relative deviations of the
   extinction cross sections calculated with the DDA and
   EMT (see Eq.~(\ref{err})) on the particle porosity.
   The particle parameters are the same as in
   Fig.~\ref{icexx} (middle panel).

\bigskip

\noindent Fig. 4.
   Porosity dependence of the normalized extinction cross sections
   calculated for spheres with small inclusions
   (DDA computations) and with the Lorenz-Mie theory using the three EMTs
   and two values of $m_{\rm compact}$.
The effect  of variations of the EMT and refractive index is illustrated.

\bigskip

\noindent Fig. 5.
   Size dependence of the extinction efficiency factors (upper panel)
   calculated for spheres with small inclusions
   (DDA computations) and with the Lorenz-Mie theory using the Bruggeman EMT.
   The porosity of the particles is ${\cal P}=0.33$.
   The effective refractive indices of the porous particles are
   indicated in Table~\ref{t-ref1}.
The lower panel shows the percent difference between
   DDA results and Bruggeman EMT calculations as defined by Eq.~(\ref{err}).
The effect of variations of the refractive indices of the inclusions is illustrated.

\bigskip

\noindent Fig. 6.
The same as in Fig.~\ref{033} but now for
porosity ${\cal P}=0.9$.

\bigskip

\noindent Fig. 7.
Size dependence of the scattering ($Q_{\rm sca}$) and absorption ($Q_{\rm abs}$)
efficiency factors,
albedo $\Lambda$ and  the asymmetry parameter $g$ for pseudospheres
with small inclusions
   using DDA computations and the Bruggeman effective model.
   The refractive indices of inclusions are $m_{\rm compact} = 1.33+0.01i$.
   The porosity of particles is ${\cal P}=0.9$.

\bigskip

\noindent Fig. 8.
   Size dependence of the extinction efficiency factors
   calculated for prolate spheroids with small inclusions using
   DDA computations and the Bruggeman effective model.
   The refractive indices of inclusions are $m_{\rm compact} = 1.33+0.01i$,
   and the porosity of particles is ${\cal P}=0.9$.
The effect  of variations of the particle shape is illustrated.

\bigskip

\noindent Fig. 9.
   Intensity and polarization of the scattered radiation
   calculated for pseudospheres with small inclusions
   (DDA computations) and effective models
    (Bruggeman--SVM computations).
   The refractive indices of the inclusions are $m_{\rm compact} = 1.33+0.01i$,
   the porosity of particles is ${\cal P}=0.9$.


\begin{thebibliography}{99}

\bibitem{mht00} M.I. Mishchenko, J. Hovenier,  and L.D. Travis, eds.,
        {\em Light Scattering by Nonspherical Particles}
        (Academic Press, San Francisco, 2000).

\bibitem{BH83}
C.F. Bohren and D.R. Huffman, {\em Absorption  and  Scattering
   of Light by Small Particles}, John Wiley, New York (1983).


\bibitem{cval00} P. Ch\'ylek, G. Videen, D.J.W.  Geldart,
     J.S. Dobbie, and H.C.W. Tso,
     ``Effective Medium Approximations for Heterogeneous Particles'',
      In {\em Light Scattering by Nonspherical Particles},
        eds. M.I.~Mishchenko {\rm et al.,} 274--308,
        Academic Press, San Francisco (2000).

\bibitem{sih99} A.H. Sihvola,  {\em Electromagnetic Mixing Formulas
        and Applications}, Institute of Electrical Engineers, Electromagnetic
        Waves Series 47, London (1999).

\bibitem{kolgust01} L. Kolokolova, and B.\AA.S. Gustafson,
``Scattering by inhomogeneous particles: microwave analog experiments and
comparison to effective medium theory'',
 J. Quant. Spectrosc. Rad. Transfer, {\bf 70}, 611--625  (2001).

\bibitem{maron05} N. Maron and O. Maron,
``On the mixing rules for astrophysical inhomogeneous grains'',
              Monthly Notices Roy. Astron. Soc., {\bf 357}, 873--880  (2005).


\bibitem{d00} B.T. Draine,
      ``The Discrete Dipole Approximation for Light
      Scattering by Irregular Targets'',
      In {\em Light Scattering by Nonspherical Particles},
        eds. M.I.~Mishchenko {\rm et al.,} 131--145,
        Academic Press, San Francisco (2000).

\bibitem{min06}
M. Min, C. Dominik, J.W. Hovenier, A. de Koter, and L.B.F.M. Waters,
``The 10 $\mu$m amorphous silicate feature of fractal aggregates
and compact particles with complex shapes'',
  Astronomy and Astrophysics, {\bf 445}, 1005--1014  (2006).

\bibitem{vih04} N.V. Voshchinnikov, V.B. Il'in, and Th. Henning,
``Modelling the Optical Properties of Composite and Porous Interstellar Grains'',
  Astronomy and Astrophysics, {\bf 429}, 371--381  (2005).

\bibitem{df03} B.T. Draine, and P.J. Flatau,
  User Guide for the Discrete Dipole Approximation Code
  DDSCAT.6.0, astro-ph/0309069, 1--46  (2003).

\bibitem{hs93}
Th. Henning, and R. Stognienko,
``Porous grains and polarization: the silicate features'',
Astronomy and Astrophysics, {\bf 280}, 609--616 (1993).

\bibitem{lr94}
K. Lumme, and J. Rahola,
``Light scattering by porous dust particles in the dicrete-dipole
approximation'',
Astrophys. J., {\bf 425}, 653--667 (1994).

\bibitem{wcms-l94} M.J. Wolff, G.C. Clayton, P.G. Martin,  and
           R.E. Schulte-Ladbeck,
``Modeling composite and fluffy grains: the effects of porosity'',
              {Astrophys. J.}, {\bf 423}, 412--425  (1994).

\bibitem{dw01} A. Doicu, Th. Wriedt, ``Equivalent refractive index of
a sphere with multiple spherical inclusions'',
               J. Opt., {\bf A3}, 204--209  (2001).

\bibitem{mal05} P. Mallet, C.A. Gu\'erin, and A. Sentenac,
``Maxwell-Garnett mixing rule in the presence of multiple
scattering: Derivation and accuracy'',
              Phys. Rev., {\bf B72}, 014205--9  (2005).


\bibitem{koc06} M. Kocifaj, M. Gangl, F. Kundrac\'ik,
H. Horvath, G. Videen,
``Simulation of the optical properties of single composite
aerosols'', Aerosol Science {\bf 37} 1683--1695 (2006).

\bibitem{gue06} Y. Gu\'eguen, M. Le Ravalec, and L. Ricard,
``Upscaling: Effective Medium Theory, Numerical Methods
and the Fractal Dream'', Pure Appl. Geophys.,  {\bf 163}
 1175--1192 (2006).


\bibitem{v02} N.V.  Voshchinnikov,
``Optics of Cosmic Dust. I'', Astrophys. \& Space Phys. Rev., {\bf 12}, 1--182
(2004).

\bibitem{heal99} Th. Henning, V.B. Il'in, N.A. Krivova, B. Michel, and N.V.
Voshchinnikov,
``WWW Database on Optical Constants for Astronomy'',
         Astronomy and Astrophysics Supplement, {\bf 136}, 405--406 (1999).

\bibitem{jetal02} C. J\"ager, V.B. Il'in, Th. Henning,
H. Mutschke, D. Fabian, D.A. Semenov, and N.V. Voshchinnikov,
``A database of optical constants of cosmic dust analogs'',
{\bf 79--80}, 765--774 (2003).

\bibitem{chch90}
H. Chang, T.T. Charalampopoulos,
``Determination of the wavelength dependence of
refractive indices of flame soot'',
Proc. Roy. Soc. London {\bf A430},  577--591, (1990).



\bibitem{vf93} N.V. Voshchinnikov and  V.G. Farafonov,
``Optical  properties of spheroidal particles'',
Astrophys. Space Sci. {\bf 204}, 19--86 (1993).

\end{thebibliography}
\end{document}